\pdfoutput=1 
\documentclass[conference]{IEEEtran}
\pdfoutput=1
\usepackage[]{times}
\usepackage{epsfig}
\usepackage{subfigure}
\usepackage{amsmath}
\usepackage{bm,amssymb}
\usepackage{graphicx}
\usepackage[top=0.5in, bottom=0.75in, left=0.75in, right=0.75in]{geometry}
\usepackage{multirow}
\usepackage{flushend}
\usepackage{url}
\usepackage{color}

\title{Options for Control of Reactive Power by Distributed Photovoltaic Generators}

\author{
\authorblockN{Petr \v{S}ulc}
\authorblockA{New Mexico Consortium,\\ Los Alamos, \\ NM
87544, USA\\ Czech Technical University\\ in Prague, \\ Czech Republic\\Email: sulcpetr@gmail.com}
\and
\authorblockN{Konstantin Turitsyn}
\authorblockA{CNLS \& Theoretical Divison\\
Los Alamos National Lab\\
Los Alamos,\\ NM 87545, USA\\
Email: turitsyn@lanl.gov}
\and
\authorblockN{Scott Backhaus}
\authorblockA{Materials, Physics\\ \& Applications Division\\
Los Alamos National Lab\\
Los Alamos,\\ NM 87545, USA\\
Email: backhaus@lanl.gov}
\and
\authorblockN{Michael Chertkov}
\authorblockA{CNLS \& Theoretical Divison\\
Los Alamos National Lab\\
Los Alamos,\\ NM 87545, USA\\
Also with NMC\\
Email: chertkov@lanl.gov}}

\begin{document}
\maketitle
\begin{abstract}
High penetration levels of distributed photovoltaic (PV) generation on an electrical distribution circuit present several challenges and opportunities for distribution utilities.  Rapidly varying irradiance conditions may cause voltage sags and swells that cannot be compensated by slowly responding utility equipment resulting in a degradation of power quality.  Although not permitted under current standards for interconnection of distributed generation, fast-reacting, VAR-capable PV inverters may provide the necessary reactive power injection or consumption to maintain voltage regulation under difficult transient conditions.  As side benefit, the control of reactive power injection at each PV inverter provides an opportunity and a new tool for distribution utilities to optimize the performance of distribution circuits, e.g. by minimizing thermal losses.  We discuss and compare via simulation various design options for control systems to manage the reactive power generated by these inverters.  An important design decision that weighs on the speed and quality of communication required is whether the control should be centralized or distributed (i.e. local).  In general, we find that local control schemes are capable for maintaining voltage within acceptable bounds.  We consider the benefits of choosing different local variables on which to control and how the control system can be continuously tuned between robust voltage control, suitable for daytime operation when circuit conditions can change rapidly, and loss minimization better suited for nighttime operation.

\end{abstract}


\section{Introduction}
\label{sec:intro}
Displacing fossil-fired generation with renewable generation has many desirable outcomes, e.g. reduction in pollution and CO$_2$ emissions, and a significant challenge, i.e. reliable delivery of electrical power of acceptable quality nearly 100\% of the time\cite{lopes2007integrating}.  The mix of renewable generation will contain many different resources including wind, concentrating solar power, and photovoltaic (PV) at the transmission-scale, but with PV as the only presently viable option at the distribution scale.  The one challenge stated above is actually a family of challenges because each of these renewable options affects reliability and power quality in different and often multiple ways.

At the transmission scale, renewable generation projects are generally large enough to warrant individual transmission interconnection studies intended uncover issues that may need to be mitigated by the renewable generation owner such as installing certain additional equipment or operating in certain ways to mitigate the problems.  In this case, the cost of mitigation is borne by the generator creating the problem.

At the distribution scale, the size of an individual PV generator is so small that the cost of an ``interconnection study'' would be prohibitive.  However, when the penetration of PV generators on any particular distribution circuit is low, the impact is quite small and present utility systems are generally unaffected.  However, at higher penetrations the net impact of many small PV generators may accumulate and affect power quality, e.g. slowly responding utility equipment (tap changers, switchable capacitors, etc.) not keeping pace with cloud-induced rapid variations of PV generation resulting in loss of voltage regulation.  Fast-response equipment could be installed to rectify the problem (e.g. a D-STATCOM \cite{moreno2007power}), but the cost is borne by the entire rate base instead of the owners of the PV generators who are benefiting from the interconnection to the distribution grid.

A potential solution to the voltage regulation problem is to tap into the latent excess PV inverter capacity to generate or consume reactive power in an attempt to control voltage.  Although not permitted by current interconnection standards \cite{1547}, changes to these standards to allow for injecting or consuming reactive power appear eminent.  Under this scheme, the burden of providing adequate reactive power compensation is again placed upon the generator seeking access to the grid.  However, many questions still remain including:
\begin{itemize}
\item How to dispatch the excess capacity to handle major changes in circuit conditions, e.g. rapid change from a net real power export to net real power import?
\item How to split the reactive compensation duty equitably between the PV generators?
\item Whether the control should be centralized (potentially vulnerable), distributed (perhaps more robust), or a combination of the two?
\item Whether centralized or distributed, what variables should be used as inputs to the control algorithm?
\end{itemize}
Despite the challenges related to accommodating PV generators, there is also an opportunity for the utility to leverage the inverters of these PV generators to enhance its own performance such as improving power quality (i.e. voltage regulation) and reducing distribution losses via optimal management of reactive power flows.  However, these should be accomplished without placing undue burdens on the PV generators by either via excessive dispatch of reactive power or by limiting PV generation.

In the remainder of this paper, we focus on these issues surrounding the integration of high penetrations of PV generation into distribution circuits.  Section~\ref{sec:O and C} discusses these issues and their impact on reactive power control systems in more detail.  In Section~\ref{sec:control}, we discuss a few control methods that have been proposed and present our own potential control method.  Section~\ref{sec:compare} compares these methods via simulations of a distribution circuit under widely varying conditions, and we draw conclusions and discuss directions for future work in Section~\ref{sec:conclusions}.

We note that there are other approaches to optimizing the dispatch of reactive power in distribution circuits for the purpose of voltage regulation and loss minimization that could be adapted to the present problem including work by Baran and Wu \cite{89BWa,89BWb,89BWc} and Baldick and Wu \cite{90BW} and also in \cite{yona2008optimal} and \cite{tani2006coordinated}.  However, these works are somewhat specialized to optimal placement, sizing, and/or control a few large sources of reactive power where the problem at hand includes many small sources.

\section{Opportunities and Challenges}
\label{sec:O and C}
\begin{figure}
\includegraphics[width=0.5\textwidth]{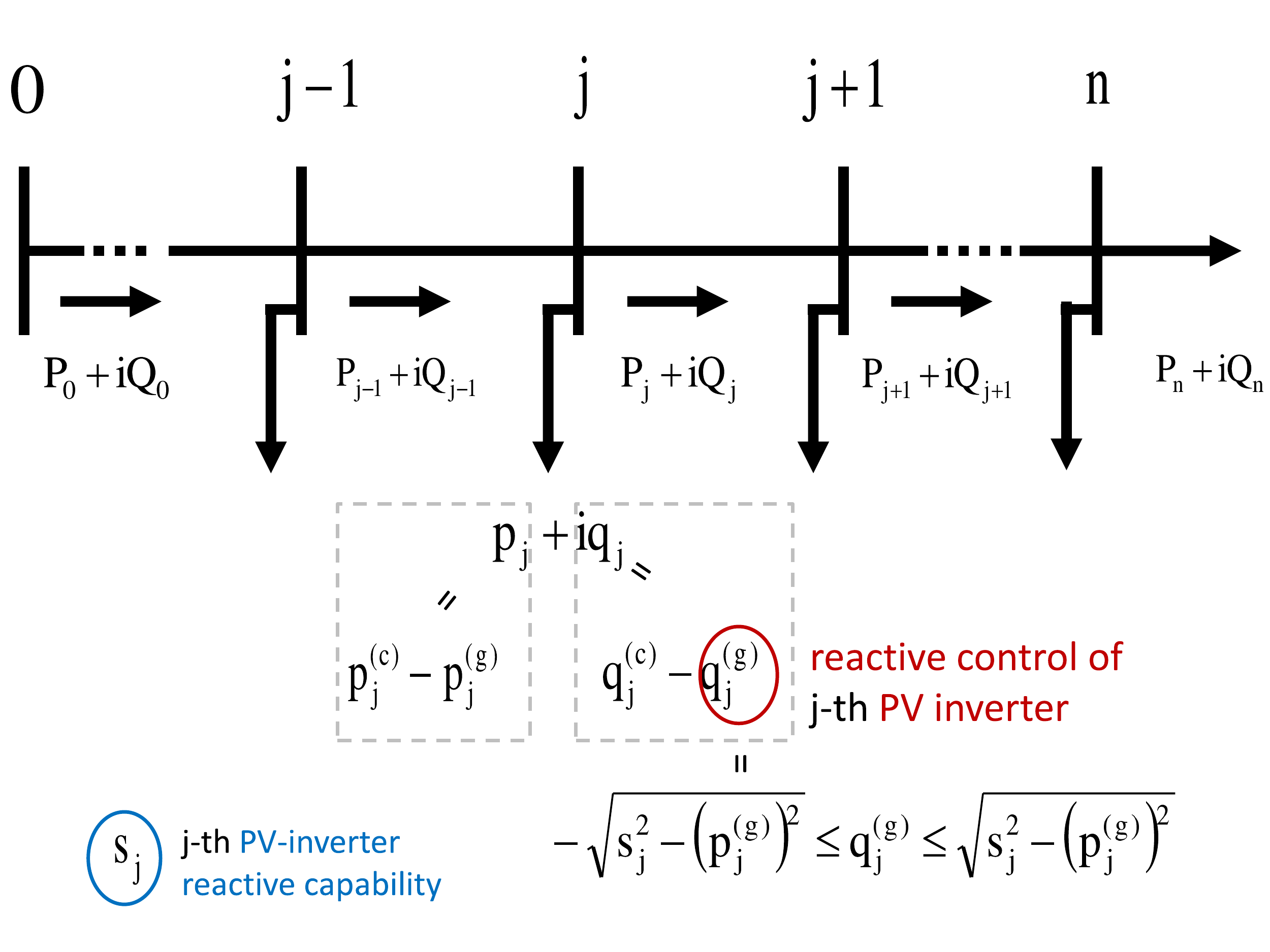}
\centering
\caption{
Diagram and notations for the radial network. $P_j$ and $Q_j$ represent real and reactive power flowing down the circuit from node $j$, where $P_0$ and $Q_0$ represent the power flow from the sub-station. $p_j$ and $q_j$ correspond to the flow of power out of the network at the node $j$, where the respective positive [negative] contributions, $p_j^{(c)}$ and $q_j^{(c)}$ [$p_j^{(g)}$ and $q_j^{(g)}$] represent consumption [generation] of power at the node. The node-local control parameter $q_j^{(g)}$ can be positive or negative but is bounded in absolute value as described in Eq.~\ref{PV_constraint}. The apparent power capability of the inverter $s_j$ is preset to a value comparable to but larger than $\max p_j^{(g)}$ .}
\label{fig:feeder}
\end{figure}

Figure~\ref{fig:feeder} introduces the schematic distribution circuit and most of the notation we will use in the remainder of the manuscript.
In our previous work \cite{KostyaMishaPetrScott,KostyaMishaPetrScott2}, we have used the \emph{LinDistFlow} equations \cite{89BWa,89BWb,89BWc} to compute voltage and power flows.  In this work, we switch to an AC solver\cite{matpower} to compute all distribution circuit quantities.  However, we introduce the problem using the \emph{LinDistFlow} equations because they provide an excellent setting for gaining intuition about the competing nature of achieving good power quality, i.e. voltage regulation, and reducing distribution circuit losses.  The \emph{LinDistFlow} equations for the circuit in Fig.~\ref{fig:feeder} can be written
\begin{eqnarray}
\label{LinDistFlow1}
P_{j+1}= P_j-p_{j+1}^{(c)}+p_{j+1}^{(g)},\\
Q_{j+1}=Q_j-q_{j+1}^{(c)}+q_{j+1}^{(g)}, \label{LinDistFlow2}\\
V_{j+1}=V_j - (r_j P_j+x_j Q_j)/V_0, \label{LinDistFlow3}
\end{eqnarray}
where $P_j+iQ_j$ is the complex power flowing away from node $j$ toward node $j+1$, $V_j$ is the voltage at node $j$, $r_j+ix_j$ is the complex impedance of the link between node $j$ and $j+1$, and $p_j+i q_j$ is the complex power extracted at node $j$. Both $p_j$ and $q_j$ are composed of local consumption minus local generation due to the PV inverter, i.e. $p_j=p_j^{(c)}-p_j^{(g)}$ and $q_j=q_j^{(c)}-q_j^{(g)}$.  Of the four contributions to $p_j+i q_j$, $p_j^{(c)}$, $p_j^{(g)}$ and, $q_j^{(c)}$ are uncontrolled (i.e. driven by consumer load and instantaneous PV generation), while the reactive power generated by the PV inverter, $q_j^{(g)}$, can be adjusted and be made either positive or negative.  As described in Section~\ref{sec:control}, $q_j^{(g)}$ is limited by the apparent power capability of the inverter $s_j$:
\begin{eqnarray}
\forall j=1,\cdots,n:\quad
\left| q_j^{(g)} \right| \leq \sqrt{s_j^2-(p_j^{(g)})^2} \equiv q_j^{max}. \label{PV_constraint}
\end{eqnarray}
Note that reactive power generation is possible only at the nodes with PV generation. For the other nodes, we take $s_j=p_j^{(g)}=q_j^{(g)}=0$.

Within the framework of \emph{LinDistFlow} equations, the rate of energy dissipation ${\cal L}_j$ and the change in voltage $\Delta V_j$ between nodes $j$ and $j+1$ of the distribution circuit are given by
\begin{eqnarray}
{\cal L}_j&=&r_j\frac{P_j^2+Q_j^2}{V_0^2},\label{lossij}\\
\Delta V_j&=&-(r_j P_j+x_j Q_j)\label{dVij}
\end{eqnarray}

\subsection{Distribution Loss Reduction vs Power Quality}
Equations~(\ref{lossij}) and~(\ref{dVij}) can be used to discuss many of the issues surrounding how to construct a control scheme to use the latent reactive power capability of PV inverters to maintain power quality and reduce losses.  First, Eq.~(\ref{lossij}) shows that losses in any circuit segment $j$ are minimized when $Q_j=0$.  However, to minimize the voltage variation, Eq.~(\ref{dVij}) would prefer if $Q_j=-(r_j/x_j)P_j$ in clear competition with loss minimization.  Therefore, in general, we should not expect a control algorithm to simultaneously provide optimal voltage regulation and minimize losses.  The trade between these two desired outcomes must be left up to engineering judgement.  However, a control scheme should be adaptable to easily allow for smooth transitions between emphasis on power quality or distribution losses.

Equation~(\ref{dVij}) also demonstrates the importance of controlling $q_j^{(g)}$ in a high PV penetration distribution circuit.  As irradiance conditions change due to cloud passage and the $p^{(g)}_j$ change rapidly, the segment flows $P_j$ can undergo rapid reversals.  A distribution circuit that was experiencing an acceptable $0.05\;p.u.$ voltage drop without PV generation could see rapid switching between the original voltage drop and a $0.05\;p.u.$ voltage rise potentially causing voltage excursions beyond acceptable bounds.  However, if the $Q_j$ can also be rapidly modified through the $q^{(g)}_j$, then the voltage variation can be controlled to within acceptable bounds.

\subsection{Centralized versus Local Control}
Equations~(\ref{lossij}) and~(\ref{dVij}) also demonstrate the complexity of developing a centralized control scheme versus local control.  The losses and voltage drop in circuit segment $j$ depend upon the flows in segment $j$, i.e. $P_j$ and $Q_j$.  Although not currently available to utilities, a centralized controller could infer the flows in each segment from smart meter data that provides $q_j$ and $p_j$ for each consumer.  With $P_j$ and $Q_j$, a centralized controller could determine the dispatch of $q^{(g)}_j$ by optimizing an objective function that includes weighted measures of losses and voltage deviations.  For this type of centralized control, the communication requirements and additional system vulnerability due to reliance on communication may outweigh the potential performance benefits.  In addition, latency in communication and control may degrade performance during rapid changes in cloud cover.

Local control schemes that act on local variables will not suffer from latency and are much less vulnerable as they do not depend upon communication for their operation (limited communication may be employed by a utility to change control algorithms perhaps up to several times during the day as overall circuit conditions change \cite{EPRI2010}).  However, truly local schemes will only have access to local flows $p_j$ and $q_j$ and, without access to the segment flows $P_j$ and $Q_j$, cannot guarantee optimal control.  This suggests a local scheme must rely upon heuristics to infer enough information about $P_j$ and $Q_j$ to take appropriate control actions.

In recent work, \cite{KostyaMishaPetrScott} we have compared centralized and local approaches to the control of reactive power. We have shown that, for a realistic distribution circuit, a local control scheme that simply supplies the local reactive power consumption (i.e. $q^{(g)}_j=q^{(c)}_j$) can achieve almost 80\% of savings in losses when compared to a centralized control based on solving the full optimization problem. Losses were actually reduced farther by blending in another heuristic to infer $P_j$ and $Q_j$ to reduce voltage drops\cite{KostyaMishaPetrScott2}.  The additional heuristic works well in reasonably high PV penetration scenarios when the circuit is importing or exporting power.  However, when PV generation and load on the circuit are in relatively close balance, the heuristic breaks down and may actually result in reduced performance.  When a circuit is in balance, the $P_j$ randomly change sign from segment to segment, and the need to dispatch $q^{(g)}_j$ to regulate voltage is not high.  Considering the advantages in speed and reliability of local versus centralized control and the comparable performance we have simulated in previous work\cite{KostyaMishaPetrScott,KostyaMishaPetrScott2}, we only consider local control in the remainder of this manuscript.

\subsection{Equitable Treatment of PV Generators}
Dispatching $q^{(g)}_j$ places additional duty on the inverters of individual PV generators which may lead to reduced lifetime and increased lifecycle cost.  Reference~\cite{EPRI2010} seeks to equitably divide the reactive power duty by setting the maximum positive and negative $q^{(g)}_j$ dispatch proportional to the capacity of the inverter and PV generator.  However, the variable that controls $q^{(g)}_j$ between these two extremes is the local voltage $V_j$.  Therefore, PV generators that are located on a distribution circuit where the voltage is continually above or below $1\;p.u$. will have to endure extra duty compared to those located where the voltage is usually close to $1\;p.u.$  Retail customers typically have no choice where they are located along a circuit.  In addition, their location relative to a substation may change from day to day depending on how the entire distribution system is configured.  Therefore, customers should not be penalized based on this location.  In one the alternative control schemes presented in this manuscript, we base control of $q^{(g)}_j$ solely on $p^{(c)}_j$, $p^{(g)}_j$, and $q^{(c)}_j$ with reactive power limits set by the capacity of the inverter so that $q^{(g)}_j$ does not depend on location along a circuit.

In this manuscript, we only discuss the dispatch of reactive power and do not consider the calls for limiting PV generation.  Although we have not encountered situations where this control action is required, Ref.~\cite{EPRI2010} provides a framework for an equitable division of generation reductions.

\section{Modeling Details}
\label{sec:model details}

\subsection{Inverter model}
\begin{figure}
\includegraphics[width=0.5\textwidth]{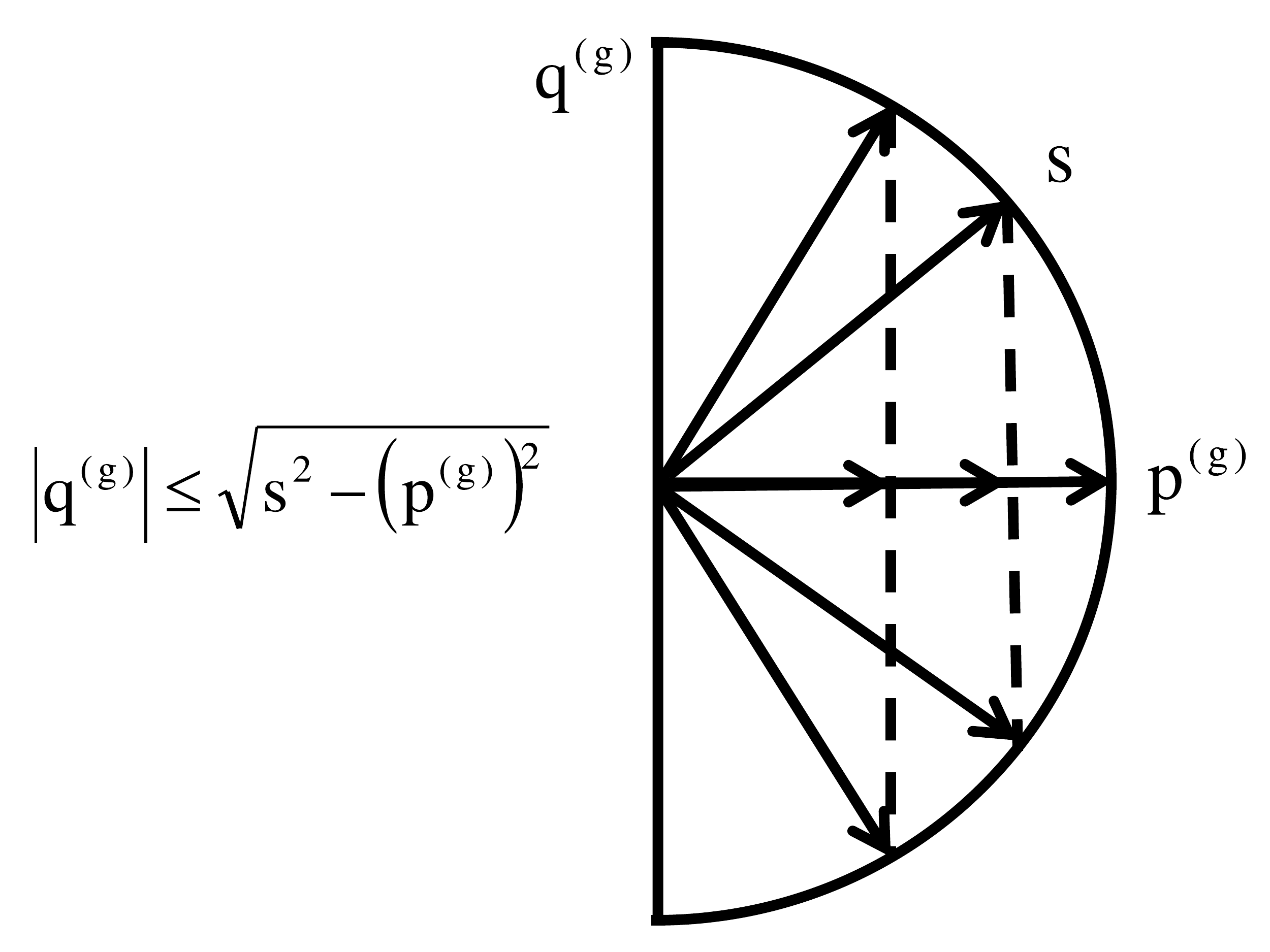}
\centering \caption{When $s$ is larger than $p^{(g)}$, the inverter can supply or consume reactive power $q^{(g)}$.  The inverter can dispatch $q^{(g)}$ quickly (on the cycle-to-cycle time scale) providing a mechanism for rapid voltage regulation.  As the output of the PV panel array $p^{(g)}$ approaches $s$, the range of available $q^{(g)}$ decreases to zero.
}
\label{fig:complex}
\end{figure}

An inverter attached to a PV generator is not an infinite source or sink of reactive power.  Its instantaneous reactive power capability is limited by its fixed apparent power capability $s_j$ and the variable real power generation $p^{(g)}_j$.  To describe this limitation mathematically, we adopt a model of PV inverters previously described in \cite{KostyaMishaPetrScott} and \cite{08LB} where the range of allowable reactive power generation is given by $|q^{(g)}|\leq \sqrt{s^2-(p^{(g)})^2}\equiv q^{max}$.  This relationship is also described by the phasor diagram in Fig. \ref{fig:complex}.  On a clear day with the sun angle aligned with the PV array, $p^{(g)}=p^{(g)}_{max}$ and the range of available $q^{(g)}$ is at a minimum.  Although $s_j$ relative to $p^{(g)}_{max}$ could be treated as a free parameter subject to optimization, our previous work \cite{KostyaMishaPetrScott} found that $s_j \approx 1.1\;p^{(g)}_{max}$ provides enough freedom in $q^{(g)}_j$ to realize the majority of the reduction in distribution losses.  Under these conditions, $|q^{(g)}_j| \leq 0.45\;p^{(g)}_{max}$ when $p^{(g)}_j=p^{(g)}_{max}$.  The choice of $s_j \approx 1.1\;p^{(g)}_{max}$ seems reasonable because inverters are available in discrete sizes and inverters are likely oversized somewhat compared to $p^{(g)}_{max}$

\subsection{Description of the prototypical distribution circuit}
\label{sec:rural}
The configuration of the distribution circuit model we consider is similar to one we previously used\cite{KostyaMishaPetrScott,KostyaMishaPetrScott2}.  Many of the circuit parameters are based on one of the 24 prototypical distribution circuits described in \cite{08SCCPET}; the nominal phase-to-neutral voltage $V_0$ is $7.2 kV$, line impedance is $(0.33+0.38i) \Omega /km$ and constant along the circuit, and the distance between neighboring nodes is $0.2$ kilometers.  The circuit consists of $250$ nodes, and we study one level of PV penetration, i.e. $50\%$ of the nodes include PV generation.  The capacity of the inverter at each PV-enabled node is set to $s_j=2.2\;kVA$, and the maximum generation capacity is set to $p^{(g)}_{max}=2.0\;kW$.  A uniform level of maximum PV power generation assumes identical installations at each PV-enabled node (the same $p^{(g)}_{max}$ installed in the same way) and spatially uniform solar irradiance.

We consider two different load/generation cases; undergenerated and overgenerated.  The undergenerated case corresponds to a situation when there is heavy cloud cover over the entire circuit and all of the $p^{(g)}_j=0$.  The load at each node is selected from a uniform distribution between $0$ and $2.5\;kW$ giving an average net real power import per node of $1.25\;kW$.  The overgenerated case corresponds to a clear sky where all $p^{(g)}_j=2\;kW$.  The load at each node is again selected from a uniform distribution, but the limits are now $0$ and $1\;kW$.  The average generation per node is then $1\;kW$ and the average load per node is $0.5\;kW$ giving an average net real power export per node of $500\;W$.  The reactive power consumed by the loads at each node, $q_j^{(c)}$, is randomly selected from uniform distribution between $0.2 p_j^{(c)}$ and $0.3 p_j^{(c)}$ corresponding power factors in the range 0.955-0.98 which is representative of residential loading\cite{kundar}.

The two cases we consider correspond to widely varying irradiance and power flow conditions.  For a given control scheme, the differences between these cases probes the robustness of the scheme to rapidly changing irradiance conditions.  To gauge the sensitivity of the control schemes considered in the manuscript to changes in circuit configurations, we consider many different realizations of a circuit.  In each realization, the $50\%$ of the nodes that are PV-enabled are selected randomly and the $p_j^{(c)}$ and $q_j^{(c)}$ distributions are sampled each time.

\subsection{Power Flow Solution Method}
\label{sec:powerflow}
We use the Matpower package\cite{matpower} to solve the AC power flow equations.
The package implements the Newton-Raphson method to solve the equations
\begin{eqnarray}
0 &=& - (p^c_i - p^g_i) + \sum_{k=1}^N |V_i| |V_k| ( G_{ik} \cos \theta_ik + B_{ik} \sin \theta_{ik}) \\
0 &=& - (q^c_i - q^g_i) + \sum_{k=1}^N |V_i| |V_k| ( G_{ik} \sin \theta_ik - B_{ik} \cos \theta_{ik}),
\end{eqnarray}
where $G_{ik}$ and $B_{ik}$ are the real and imaginary parts of the impedance matrix $Y$, respectively, and $\theta_{ik}$ is the difference in voltage phase angle between buses $i$ and $k$.

The Matpower package solver does not support situations where $q^{(g)}_i$ becomes a function of voltage.  For the control schemes that require this (described below), we modify the implementation of the Newton-Raphson method to take into account the variation of $q^{(g)}_i$ with respect to the voltage.

\section{Control Schemes}
\label{sec:control}
\subsection{Control on Local Voltage Only}
\begin{figure}
\includegraphics[width=0.5\textwidth]{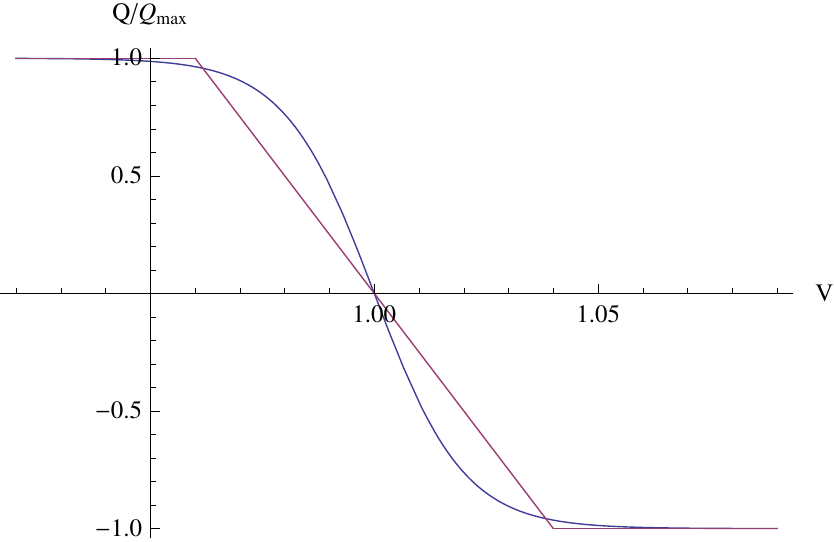}
\centering
\caption{Dark red, piecewise linear curve: A simplified version of the proposed $q^{(g)}_j$ control function from Ref.~\cite{EPRI2010}.  Blue smooth curve:  Equation~(\ref{EPRI control}) with $\delta=0.04$ which we use here to improve the convergence properties of the AC solver while closely representing the control function in Ref.~\cite{EPRI2010}. }
\label{EPRI}
\end{figure}

Reference~\cite{EPRI2010} has proposed a reasonable framework for local control of reactive power produced by the inverters of PV generators.  Although four different modes of control are proposed, each consists of a set of piecewise linear relationships between $q^{(g)}_j$ and $V_j$.  A simplified version of mode `PV1' from Ref.~\cite{EPRI2010} is shown in Fig.~\ref{EPRI} Although not specified in Ref.~\cite{EPRI2010}, we take $q^{(g)}_j=0$ at $V_j=1\;p.u.$, and we take the saturated values of $q^{(g)}_j$ at high and low values of $V_j$ to be given by $q_j^{max}$ defined in the earlier discussion of the inverter model.  Except for the dynamic definitions of $q_j^{max}$, this is essentially a proportional control scheme where $q^{(g)}_j$ depends linearly on $V_j$.

In our attempts to utilize this scheme, the Matpower AC solver\cite{matpower} we employed showed convergence problems which we diagnosed as the solution jumping back and forth across the points of discontinuous first derivative.  To ease this difficulty, we smoothed the control function in Fig.~\ref{EPRI} using a sigmoid function, i.e.
\begin{equation}\label{EPRI control}
G(q_j^{max},V_j,\delta)= q_j^{max}\left(1-\frac{2}{1+\exp[-4(V_j-1)/\delta]}\right).
\end{equation}
Here, $\delta$ is simply a parameter that controls how closely the smoothed control function approximates the sharp transitions of the original control function.  In this work we have taken $\delta=0.04$, and $G(q_j^{max},V,0.04)$ is plotted in Fig.~\ref{EPRI} for comparison to the piecewise continuous control proposed in Ref.~\cite{EPRI2010}.

\subsection{Control on Local Flows Only}
\label{sec:Local Flows}
The control function $G(q_j^{max},V,\delta)$ describes a form of local control that only depends on the voltage at the point of inverter connection.  Since this voltage is immediately available to the inverter, $G$ describes a scheme that would be very convenient to implement.  However, if the predicted replacement of mechanical meters with smart meters occurs, information in addition to voltage may be available to control an inverter's $q^{(g)}_j$.  We assume a smart meter will be able to provide both real and reactive net power flows and that these can be communicated to the local PV inverter.  The inverter will already have measures of its own real and reactive power generation.  The combination of this data will easily provide the inverter with near real-time access to the three local, uncontrolled power flows, i.e. $p^{(c)}_j$, $q^{(c)}_j$, and $p^{(g)}_j$.  It is from these local power flows, as opposed to $V_j$, that we construct an alternative control scheme.  We could also explicitly include $V_j$, however as discussed earlier, this choice could easily lead to inequities based upon where a PV generator is located along a distribution circuit.

In previous work \cite{KostyaMishaPetrScott2}, we have analyzed control schemes of the general form
\begin{equation}\label{control}
	q_j^{(g)} = F_k(p_j^{(g)}, p_j^{(c)}, q_j^{(c)}),
\end{equation}
and consistent with constraint (\ref{PV_constraint}).  Here, we summarize some of that work.  The control scheme is local in that $q_j^{(g)}$ depends only on $p_j^{(g)}, p_j^{(c)}, q_j^{(c)}$.  Similar to the voltage scheme discussed above, we also assume that the control is homogeneous over the line: all inverters are programmed in the same way, and explicit dependence on the bus number $j$ enters through the inverter's dynamically-determined capability $q_j^{max}$ which in turn depends on $s_j$ and $p^{(g)}_j$ through Eq.~(\ref{PV_constraint}).

It is useful to introduce the following ``helper" function, $\text{Constr}_{j}$, meant to enforce the constraint \eqref{PV_constraint}:
\begin{equation}\label{multi}
 \text{Constr}_{j}[q] = \begin{cases}
q, &    \left| q \right|  \leq  q_j^{max}  \\
(q/|q|)q_j^{max}, & \text{otherwise}
\end{cases}
\end{equation}
A local control scheme proposed in \cite{KostyaMishaPetrScott} was based on the heuristic that losses are minimized when the reactive flows $Q_k$ are zero, and the $q_j^{(g)}$ were chosen to minimize the {\it net} reactive power consumption $q_j^{(c)}-q_j^{(g)}$ at each node:
\begin{equation} \label{losscontrol}
	F_k^{(L)} = \text{Constr}_{k}[q_k^{(c)}].
\end{equation}
In this scheme, the inverter supplies the local consumption of reactive power up to the limits imposed by its capacity $s_j$ and generation $p^{(g)}_j$.  This scheme was shown to be very effective in reducing the losses.  However, as discussed in Section~\ref{sec:O and C}, loss minimization and voltage regulation are competing objectives and minimizing losses does not ensure voltage regulation.

In Ref.~\cite{KostyaMishaPetrScott2}, the control in Eq.~(\ref{losscontrol}) was extended to consider voltage regulation.  Equation~(\ref{LinDistFlow3}) suggests that, to reduce variations in $V_j$, we should minimize the absolute value of the combined power flow $r_j P_j + x_j Q_j$. Note that for many circuits, the ratio of $r_j/x_j = \alpha$ is nearly constant for all $k$ and depends only on the configuration and size of the conductors used. Thus the absolute value of $r_j P_j + x_j Q_j$ will be exactly zero if for every load node we ensure that $p_j^{(c)} - p_j^{(g)} + \alpha \left( q_j^{(c)} - q_j^{(g)} \right) = 0$ suggesting a control function $F_j^{(V)}$ aimed at minimizing  voltage variations without regard for losses:
\begin{equation}\label{voltcontrol}
	F_j^{(V)} = \text{Constr}_{k}\left[  q_j^{(c)} + \frac{p_j^{(c)} - p_j^{(g)}}{\alpha} \right].
\end{equation}

The control in Eq.~(\ref{losscontrol}) seeks to minimize losses while Eq.~(\ref{voltcontrol}) seeks to regulate voltage.  A continuous compromise between the two objectives in can be achieved via the following nonlinear combination
\begin{equation}\label{hybrid}
	F_j(K) = \text{Constr}_{k}\left[ K F_j^{(L)} + (1-K) F_j^{(V)} \right],
\end{equation}
where $K$ is a single parameter controlling the trade off between the two objectives in Eq.~(\ref{multi}). At $K=1$ we recover the loss reduction scheme of Eq.~(\ref{losscontrol}), whereas at $K=0$ we recover the voltage regulation scheme of Eq.~(\ref{voltcontrol}).  Through the parameter $K$, we now have a simple method to smoothly adapt the control scheme, if necessary, as circuit conditions change.

\subsection{Hybrid Control}
\label{sec:combined control}
We have argued that inclusion of $V_j$ as an input to the control method may result in inequitable division of reactive power generation duty.  However, heuristics used to create the control in Eq.~(\ref{hybrid}) may, under certain circumstances, fail to provide a good estimate of the segment flows $P_j$ and $Q_j$.  Without knowledge of $V_j$, the control in Eq.~(\ref{hybrid}) has no way of correcting if $V_j$ has moved significantly from $1\;p.u.$  To correct this shortcoming, we create a hybrid control that combines Eqs.~(\ref{EPRI control}) and~(\ref{hybrid}).

The concept behind the hybrid control is similar to that used in blend $F_j^{(L)}$ and $F_j^{(V)}$ in Eq.~(\ref{hybrid}).  We desire that if $V_j=1\;p.u.$, then the control of $q^{(g)}_j$ is completely governed by Eq.~(\ref{hybrid}).  However, if $V_j$ has fallen significantly below $1\;p.u.$, then $q^{(g)}_j\rightarrow q_j^{max}$.  Similarly, if $V_j$ has risen significantly above $1\;p.u.$, then $q^{(g)}_j\rightarrow -q_j^{max}$.  A simple control that achieves this behavior is given by
\begin{equation}\label{H}
H_j(K,V_j)=F_j(K)+G(q_j^{max}-F_j(K),V_j,\delta).
\end{equation}
At $V_j=1$, $G=0$ and $H_j=F_j$.  For $V_j<<1$, $G\rightarrow q_j^{max}-F_j(K)$ and $H_j\rightarrow q_j^{max}$.  Finally, if $V_j>>1$, $G\rightarrow -(q_j^{max}-F_j(K))$ and $H_j\rightarrow -(q_j^{max}+ 2 F_j(K))$ which is still bounded between $\pm q_j^{max}$.  Here, we have chosen to blend the $G$ and $F$ control in one particular way.  There are clearly other ways to achieve these, but we leave this for future study.

\section{Simulations: Results and Discussions}
\label{sec:compare}
The control schemes described in Section~\ref{sec:control} are simulated on the distribution circuit described in Section~\ref{sec:rural}.  The node voltages and the distribution circuit losses are calculated for both the under and overgenerated cases.  For the undergenerated case, the node voltages are presented in Fig.~\ref{undervolt} and the losses in Fig.~\ref{underloss}.  For the overgenerated case, the node voltages are presented in Fig.~\ref{overvolt} and the losses in Fig.~\ref{overloss}.

\subsection{Base Case-No Control with $q^{(g)}_j=0$}
The base case where all $q^{(g)}_j=0$ corresponds to the situation imposed by the current distributed generation interconnection standards \cite{1547}.  In the undergenerated case where $P_j$ and $Q_j$ are in the same direction, the voltage deviation below $1\;p.u.$ is quite large at about $0.07\;p.u.$  In the overgenerated case, $P_j$ and $Q_j$ are now in opposite directions for the majority of $j$ and, in spite of not taking any actions, the maximum voltage rise of about $0.015\;p.u.$ is relatively small.  In both cases, the maximum deviations take place at or very near to the end of the distribution circuit. However, during party cloudy daylight hours, the voltage will swing $0.085\;p.u.$ as the circuit transitions between the under and overgenerated cases we consider--uncomfortably close to allowable limits.  Under higher load or PV generation conditions, the voltage swings would easily exceed $0.1\;p.u.$ demonstrating the need for control of reactive power in high PV penetration scenarios.  In the rest of the discussion, we use the losses incurred in this base case to normalize the losses for the other control schemes.

\subsection{Control on Local Voltage Only-$G(V)$}
Controlling the $q^{(g)}_j$ on local voltage via Eq.~(\ref{EPRI control}) provides excellent voltage regulation with an approximate drop of $0.027\;p.u.$ in the undergenerated case and a $0.008\;p.u.$ rise in the overgenerated case.  The total voltage swing on a partly cloudy day is reduced to about $0.035\;p.u.$--a significant improvement over situation when $q^{(g)}_j=0$.  However, we note that the relative losses are increased by about $5\%$ in the undergenerated case and by $20\%$ in the overgenerated case.  The significant increase in the overgenerated case can be traced to the rise of $V_j$ over $1\;p.u.$ which forces the inverters to consume reactive power increasing the flows $Q_j$ and the dissipation.  The large increase in $Q_j$ is in part driven by our choice of $q^{(g)}_j\rightarrow \pm q^{max}_j$ as $V_j$ deviates significantly from $1\;p.u.$ By reducing these limits, we could reduce the dissipation, but voltage regulation will deteriorate as the control scheme would begin to resemble $q^{(g)}_j=0$.  In a comparison of all the schemes (discussed below) and in Fig.~\ref{all}, we show how reducing the $q^{(g)}_j$ limits impacts performance.

\subsection{Control on Local Flows Only-$F(K)$}
At $K=0$, this scheme emphasizes voltage regulation through Eq.~(\ref{voltcontrol}).  Therefore, it is not surprising that near $K=0$ this scheme has similar voltage regulation performance at $G(V)$.  Near $K=1$ where loss reduction is emphasized via Eq.~(\ref{losscontrol}), $F(K)$ has significantly less dissipation than $G(V)$, but the voltage regulation is nearly as poor as $q^{(g)}_j=0$.  Clearly, there is no globally optimum value of $K$ because, as we have discussed relative to Eqs.~(\ref{lossij}) and~(\ref{dVij}), voltage regulation and loss reduction are in competition.  The choice of $K$ is then left up to engineering judgement and Fig.~\ref{all} (discussed below) provides a useful guide.

\subsection{Hybrid Control-$H(K,V)$}
Hybrid control via $H(K,V)$ attempts to contain large voltage deviations by smoothly switching from $F(K)$ to $G(K)$, i.e. better voltage control, as $V_j$ move significantly away from $1\;p.u.$  However, if $V_j$ is close to $1\;p.u.$, $H(K,V)$ looks more like $F(K)$ which allows for a greater emphasis on loss reduction.  For both the over and undergenerated cases, the blending of $G(K)$ with $F(K)$ works well for voltage regulation with $H(K,V)$ outperforming both $G(K)$ and $F(K)$ for $K<1$.  For losses, the picture is not as clear.  In the undergenerated case, $H(K,V)$ for $K<1$ still results in increased losses over the base case and greater losses than for $G(V)$.  In contrast, $H(K,V)$ for the overgenerated case results is significant loss reductions over the base case and $G(K)$ around $K=1$.  The choice of $K$ for this control scheme is again left up to engineering judgement.

\begin{figure}
\includegraphics[width=0.5\textwidth]{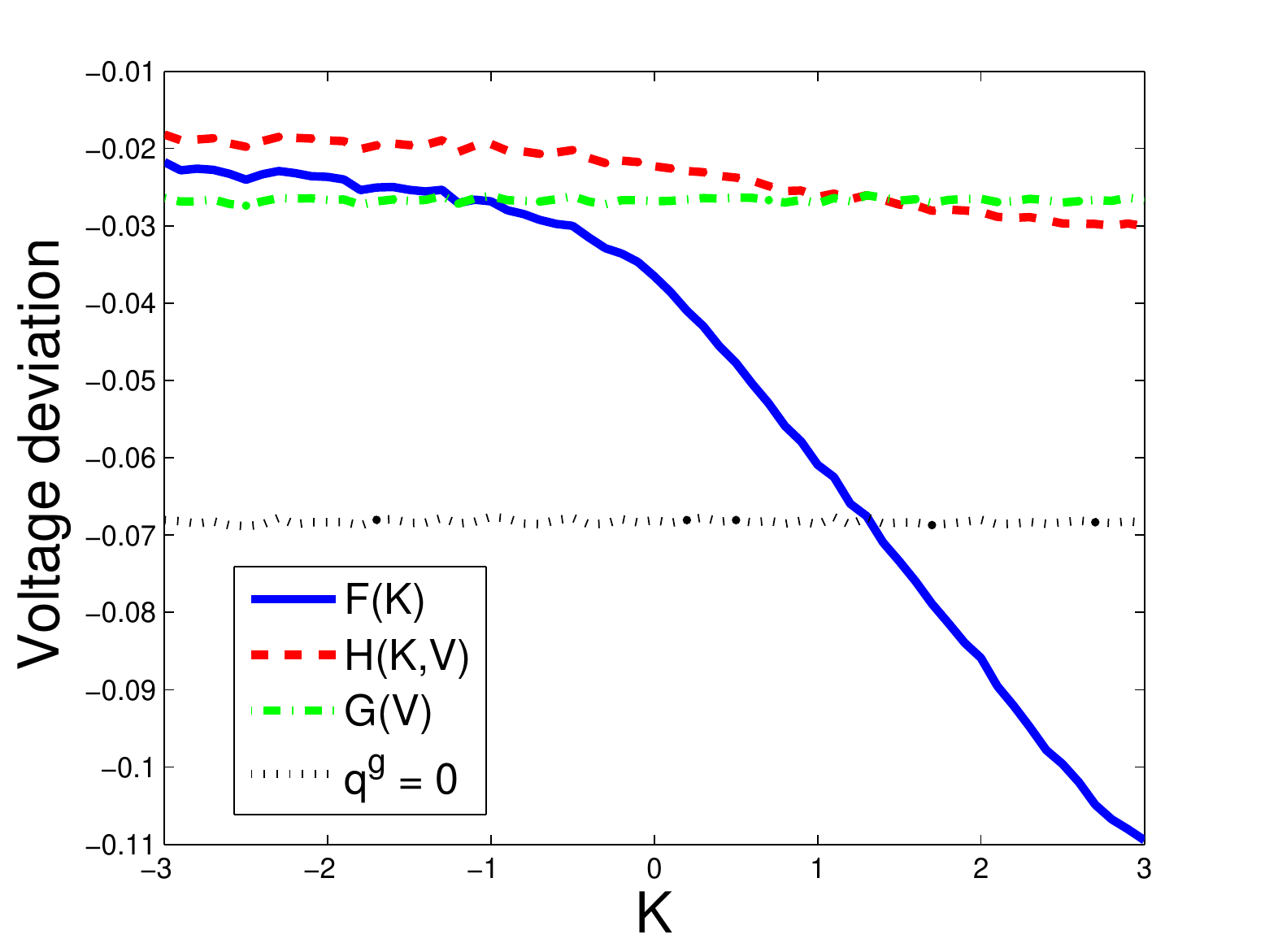}
\centering
\caption{Undergenerated case: Maximum deviation of $V_j$ from $1\;p.u.$.  Black dotted line: No control,$q^{(g)}_j=0$; green dashed line: control via local voltage, blue solid line: control via local power flows, red dashed line: hybrid control. }
\label{undervolt}
\end{figure}

\begin{figure}
\vspace{-1.4 in}
\includegraphics[width=0.5\textwidth]{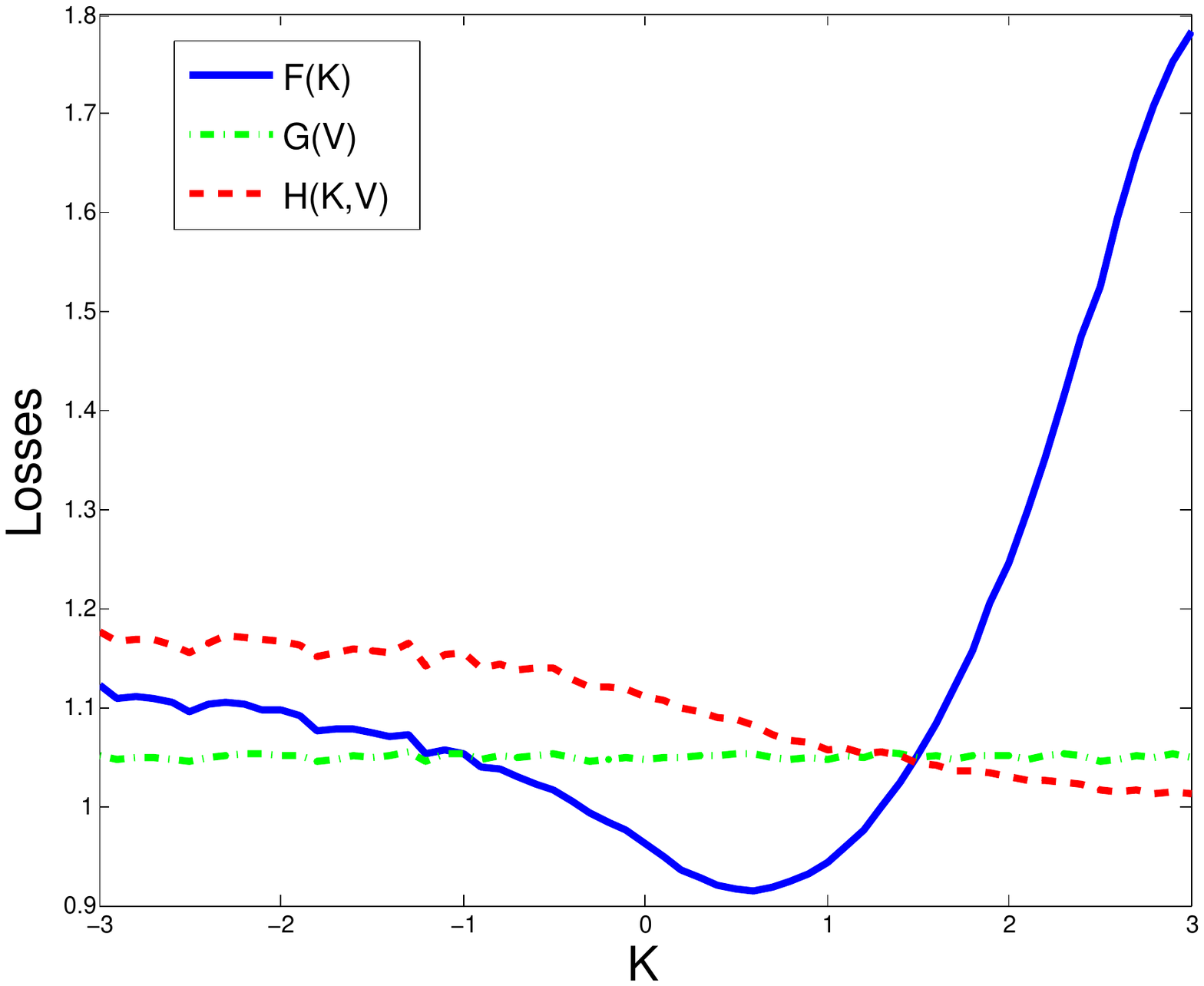}
\vspace{-1.4 in}
\centering
\caption{Undergenerated case: Distribution circuit losses normalized by the losses when $q^{(g)}_j=0$.  Lines are the same as in Fig.~\ref{undervolt}.}
\label{underloss}
\end{figure}

\begin{figure}
\vspace{-.2 in}
\includegraphics[width=0.5\textwidth]{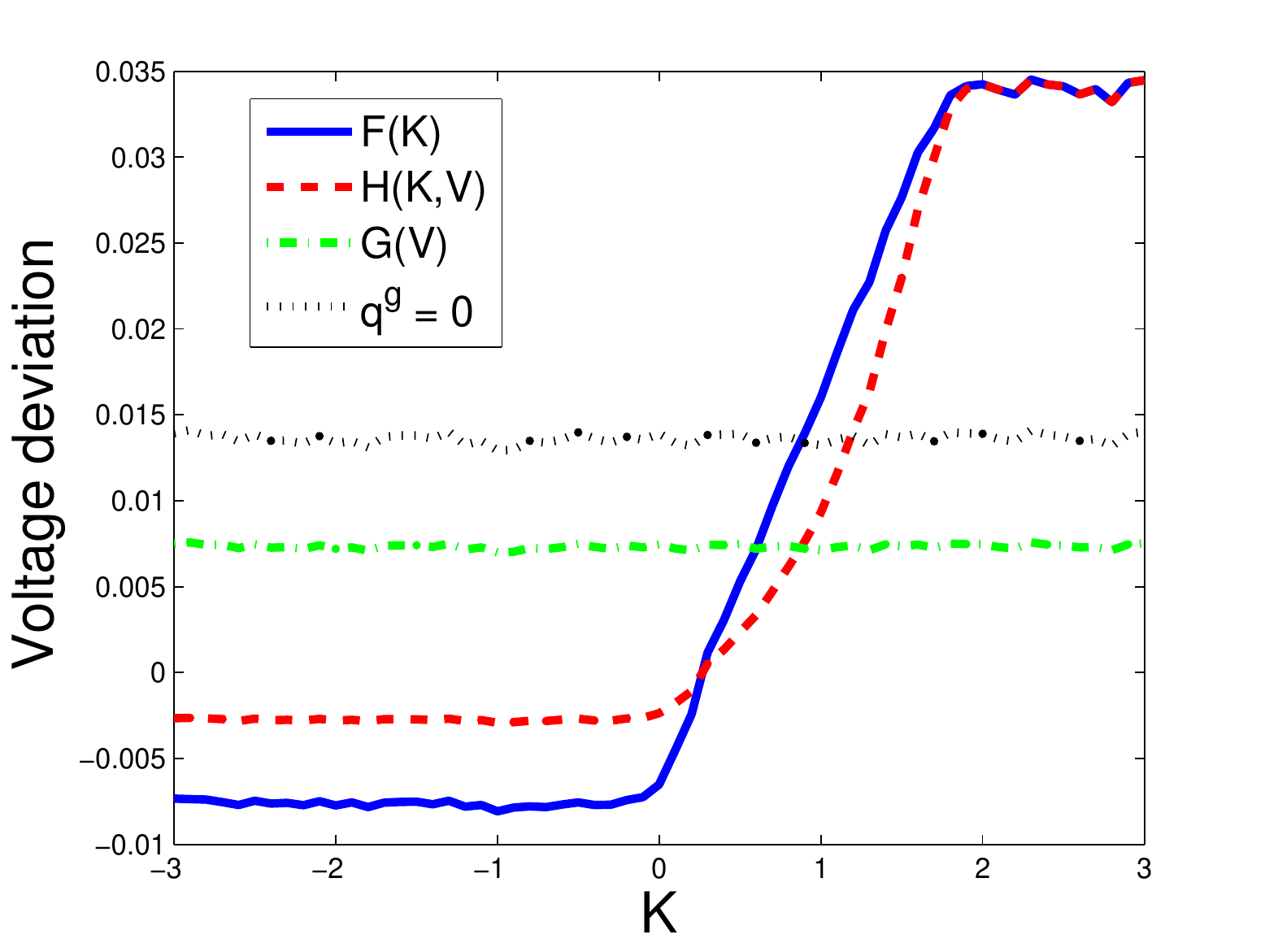}
\centering
\caption{Overgenerated case: Maximum deviation of $V_j$ from $1\;p.u.$.  Lines same as in Fig.~\ref{undervolt}.}
\label{overvolt}
\end{figure}

\begin{figure}
\vspace{-1.1 in}
\includegraphics[width=0.5\textwidth]{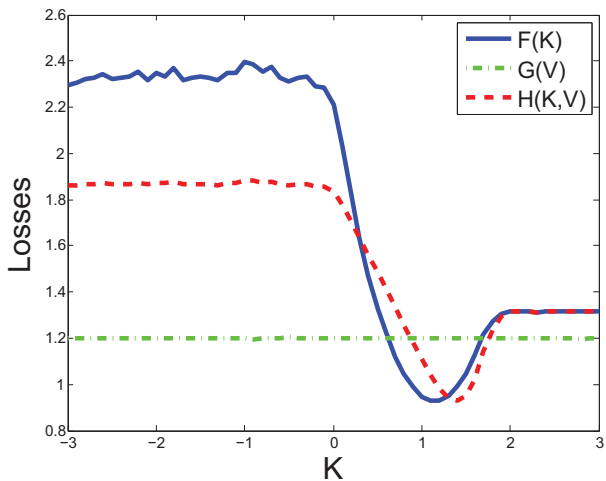}
\vspace{-1.1 in}
\centering
\caption{Overgenerated case: Distribution circuit losses normalized by the losses when $q^{(g)}_j=0$.  Lines same as in Fig.~\ref{underloss}.}
\label{overloss}
\end{figure}

\subsection{Comparison of Control Schemes}
\begin{figure}
\vspace{-1.1 in}
\includegraphics[width=0.5\textwidth]{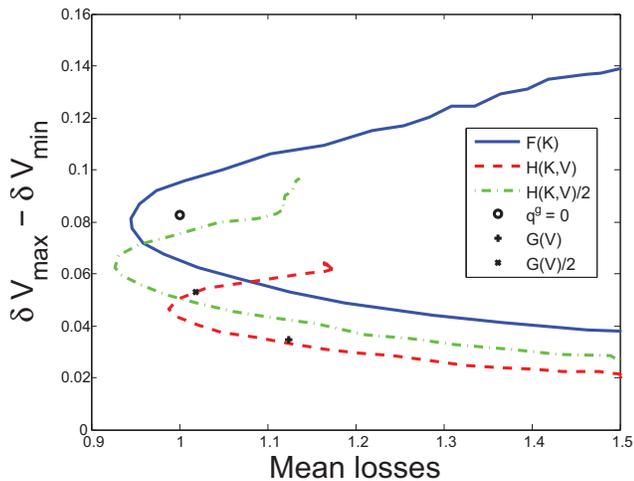}
\vspace{-1.1 in}
\centering
\caption{Maximum per unit voltage swing experienced on the distribution circuit as the loading conditions transition from the over to undergenerated case versus the average relative losses in the over and undergenerated cases.  }
\label{all}
\end{figure}

Comparison of the different control schemes is difficult due to a lack of a global optimum in both voltage regulation and loss minimization--even for a single scheme for a single case (i.e. over or undergenerated).  In this work, we are interested in finding a robust control scheme that can handle the rapid variations in power flows as a circuit with a high penetration of PV undergoes rapid changes in solar irradiance.  Therefore, we collapse the over and undergenerated results into a single plot by computing the maximum voltage swing experienced during the transition from over to undergenerated, i.e. the voltages in Fig.~\ref{overvolt} minus the voltages in Fig.~\ref{undervolt}.  These values make up the vertical axis in Fig.~\ref{all}.  To compare losses, we simply average the relative losses from Figs.~\ref{underloss} and~\ref{overloss} and these make up the horizontal axis in Fig.~\ref{all}.  Clearly, different weightings of the over and undergenerated are possible, as are other constructions.  However, for this work, we simply choose the average.

In Fig.~\ref{all}, the points for $G(V)$ and $q^{(g)}=0$ are two ends of what should be a continuous smooth curve because, as the $q^{(g)}_j$ limits are decreased, $G(V)\rightarrow q^{(g)}=0$.  The point labeled $G(V)/2$ in Fig.~\ref{all} was generated with the limits reduced by one half and will also lie on this smooth curve.  The curve for the control $F(K)$ generally lies higher and to the right of the points for $G(K)$ (and its scaled versions) showing that $F(K)$ generally gives both poorer voltage regulation and higher losses.  The $F(K)$ curve does fall lower and to the left of the points for $G(K)$ (and its scaled versions) for some values of $K$, but the voltage deviations in those cases are already approaching $0.1\;p.u.$

The hybrid control $H(K,V)$ and a scaled version labeled $H(K,V)/2$ (where we have reduced the upper and lower $q^{(g)}_j$ limits by a factor of 2) generally lie below and to the left of $G(K)$ and its scaled versions.  The implication is that inclusion of local real and reactive power flows into a high PV penetration control scheme can lead to better performance than simply utilizing the local voltage.  The curves for $H(K,V)$ and $H(K,V)/2$ are each parameterized by $K$ and the family of such curves is parameterized by the scaling factor (here we restricted the scaling factors to 1 and 1/2).  Plots, similar to that in Fig.~\ref{all}, for other distribution circuits will clearly aid in the selection of $K$ and the scaling factor.

\section{Conclusions and Path Forward}
\label{sec:conclusions}
In this work, we have developed schemes for controlling PV inverter-generated reactive power for high PV penetration distribution circuits.  In addition, we have developed a method for assessing the robustness of these and other control schemes during rapid variations in solar irradiance.  Our metrics of performance included the maximum per unit voltage change experienced during the transition from an over to undergenerated load condition and the average of the dissipation for the two conditions.  We have compared the control schemes developed in this work to those proposed by others\cite{EPRI2010} and have reached several conclusions:
\begin{itemize}
    \item The fundamental competition between voltage regulation and power quality, in general, prohibits control schemes from achieving a global optimum, i.e. a minimum in voltage deviations and circuit dissipation
    \item For the cases considered, control schemes that only require access to the local variables are sufficient to provide adequate voltage regulation.
    \item The inclusion of local real and reactive power flows, in addition to local voltage, leads to better control system performance.
\end{itemize}

In this work, we have focused mainly on the rapid transitions in loading that a high PV penetration circuit can experience during changes in solar irradiance.  However, there are still open questions related to dispatch of reactive power from the PV inverters during other times.  For instance, during nighttime hours when there is no PV generation and little concern about rapid changes in loading, is it equitable to use the reactive capability of the PV inverters to improve the circuit performance?  If so, which control scheme (among those considered here or others) provides the best performance?  We hope that this work spurs others to consider these questions in greater detail.

\section*{Acknowledgment}

We are thankful to all the participants of the ``Optimization and Control for Smart Grids" LDRD DR project at Los Alamos
and Smart Grid Seminar Series at CNLS/LANL for multiple fruitful discussions. Research at LANL was carried out under the auspices of the National Nuclear Security Administration of the U.S. Department of Energy at Los Alamos National Laboratory under Contract No. DE C52-06NA25396. P\v{S} and MC acknowledges partial support of NMC via NSF collaborative grant CCF-0829945 on ``Harnessing Statistical Physics for Computing and Communications''.

\bibliographystyle{IEEEtran}
\bibliography{SmartGrid}

\end{document}